Author version of accepted article:

Cite as

Niazi, Muaz, and Amir Hussain. "Agent-based computing from multi-agent systems to agent-based models: a visual survey." *Scientometrics* 89.2 (2011): 479.

# Agent-based computing from multi-agent systems to agent-based Models: a visual survey

Muaz Niazi[1,2] and Amir Hussain[2]

[1]*Department of Computer Science,*
*COMSATS Institute of IT,*
*Islamabad,*
*Pakistan*

[2]*Institute of Computing Science and Mathematics,*
*School of Natural Sciences,*
*University Of Stirling,*
*Scotland, UK*

Emails: muaz.niazi@gmail.com, ahu@cs.stir.ac.uk

[1]

Agent-Based Computing is a diverse research domain concerned with the building of intelligent software based on the concept of "agents". In this paper, we use Scientometric analysis to analyze all sub-domains of agent-based computing. Our data consists of 1,064 journal articles indexed in the ISI web of knowledge published during a twenty year period: 1990-2010. These were retrieved using a topic search with various keywords commonly used in sub-domains of agent-based computing. In our proposed approach, we have employed a combination of two applications for analysis, namely Network Workbench and CiteSpace - wherein Network Workbench allowed for the analysis of complex network aspects of the domain, detailed visualization-based analysis of the bibliographic data was performed using CiteSpace. Our results include the identification of the largest cluster based on keywords, the timeline of publication of index terms, the core journals and key subject categories. We also identify the core authors, top countries of origin of the manuscripts along with core research institutes. Finally, our results have interestingly revealed the strong presence of agent-based computing in a number of non-computing related scientific domains including Life Sciences, Ecological Sciences and Social Sciences.

---

## Introduction

Agent-based Computing(Jennings 1999a; Wooldridge 1998) is a large and widely spread scientific domain. An agent can range from a "software agent" or "service/daemon", which might not behave very intelligently to an intelligent agent, which is based on models of artificially intelligent behavior (Wooldridge 2009; Weiss 1998). An agent could even be a representation of an interacting social component of a large system used to explore emergent global behavior in a simulation(Gilbert and Troitzsch 2005; Muaz A Niazi and Amir Hussain 2011).

Agent design and simulation go hand in hand but in completely different ways in different sub-domains of agent-based computing. So, e.g. on one hand, there are researchers whose research goals revolve around the design of various types of agents where the role of simulation is closely linked to validation of the future operation of actual or physical agents(Bellifemine et al. 2001). On the other, there are researchers, whose goal is not agent-design but rather the agent-design is a means of developing simulations which can lead to better understanding of global or emergent phenomena associated with complex adaptive systems(Macal and North 2007; M. A. Niazi and A. Hussain 2011).

This broad base of applications of this research area thus often leads to confusions regarding the exact semantics of various terms in the literature. This is perhaps tied closely to the evolution of "agent-based computing" into a wide assortment of communities. These communities have at times, perhaps nothing other than the notion of an "agent" in common with each other.

What makes the study of this domain even harder is related closely with the keywords used by the researchers. Not only are the application domains varied, Agent-based modeling (Axelrod 1997) is at times confused with similar but actually somewhat different sub-domains such as multiagent systems (Lesser 2007) in the domain of Artificial Intelligence. While at other times, agent-based modeling is referred by completely different keywords but with synonymous meanings such as "Individual-based modeling". However, all of these eventually tie in together in the domain of agent-based computing (Wooldridge 1998; Jennings 1999b). And it is not uncommon to get unexpected results from papers where the use of the word

"agent" in completely different contexts such as in Biology or Chemistry or other domains (Snead et al. 1995; Dydenko et al. 2005) where the use is completely unrelated to agent-based computing.

While Agents and multi-agent systems from the AI perspective are not less important in any way, agent-based modeling and simulation paradigm (ABM) has even been termed a revolution in the esteemed journal Proceedings of the National Academy of Sciences (Bankes 2002). ABM has found parallel applications in numerous domains as diverse as the Social Sciences(Epstein and Axtell 1996) to Biological Sciences (Gonzalez et al. 2003; Siddiqa et al. 2009; Mukhopadhyay et al. 2010), to Environmental modeling(Muaz Niazi et al. 2010; Xiaofei 2010). It is even prevalent in the modeling of business systems(Aoyama 2010) and recently in the modeling of Computational Systems such as in Wireless Sensors (Muaz Niazi and Hussain 2010) and ad-hoc Networks (Muaz Niazi and Hussain 2009).

The goal of this paper is to use citation analysis and visualization to give a scientometric overview and survey of the diversity and spread of the domain across its various sub-domains, in general, and to identify key concepts, which are mutual to the various sub-domains, in particular. These includes identifying such visual and scientometric indicators as the identification of the core journals, key subject categories, the top and most highly cited authors, the Institutes and also the countries of manuscript origin.

The structure of the rest of the paper is as follows: First we give a brief background of Visualization techniques in Scientometrics. This is followed by data collection and methodology. Next, we present the results of the study and finally we conclude the paper.

## Background

### Scientometrics

Scientometrics (Leydesdorff 2001) is the quantitative study of scientific communication. Scientometrics requires the use of a multitude of sophisticated techniques including Citation Analysis, Social Network Analysis and other quantitative techniques for mapping and measurement of relationships and flows between people, groups, organizations, computers, research papers or any other knowledge-based entities.

**Domain Visualization**

Domain visualization is a relatively newer research front. The idea is to use information visualization to represent large amounts of data in research fronts(Chen et al. 2001). This allows the viewer to look at a large corpus and develop deeper insights based on a high level view of the map(Card et al. 1999). Visualization using various network modeling tools has been performed considerably for social network analysis of citation and other complex networks(H. White and K. McCain 1998).

Various types of scientometric analyses have been recently performed for domains such as HIV/AIDS (Pouris and Pouris 2010), Scientometrics (Hou et al. 2008), Mexican Astronomers(Sierra-Flores et al. 2009), scientific collaborations (Barabási et al. 2002) and engineers in South Africa (Sooryamoorthy 2010). Extensive work on research policy has been performed by Leydesdroff (Etzkowitz and Leydesdorff 2000). In addition, network analysis techniques have been proposed in (Park et al. 2005; Park and Leydesdorff 2008)

Recently however, with newer free tools such as the CiteSpace (Chen 2006), researchers have performed information visualization to identify various patterns in complex domains of scientific literature. Some of the recent studies in this direction include visualization of the HCI domain(Chen et al. 2006), identification of the proximity clusters in dentistry research(Sandström and Sandström 2007), visualization of the pervasive computing domain(Zhao and Wang 2010), visualization of international innovation Journals(Chun-juan et al. 2010) as well as identification of trends in the Consumer Electronics Domain(M. Niazi and A. Hussain 2011).

Scientometric studies which combine co-citation analysis with visualizations greatly enhance the utility of the study. They allow the readers to quickly delve into the deeper aspects of the analysis. Co-citation analysis deals with the measurement of proximity of documents in various scientific categories and domain. These analyses include primarily the author co-citation analysis and the journal co-citation analysis. Here, the Journal co-citation analysis identifies the role of the primary journals in the domain. In contrast, the author co-citation analysis(White and Griffith 1981) especially by using visualization(H. D. White and K. W. McCain 1998) offers a view of the structures inside a domain. The idea is that the more frequently two authors are cited together, the closer the scientific relation between them. Whereas document co-citation maps co-citations of individual documents (Small 1973; Small and Griffith 1974; Griffith et al. 1974), author co-citation focuses on body of literature of individual authors. In addition, co-citation cluster analysis of documents is useful for the display of macro-structural scientific knowledge evolution. In initial cluster analysis, techniques involved clustering highly cited documents by single-link clustering and then clustering the resultant clusters up to four

times(Small 1993). Recent techniques involve the use of computational algorithms to perform this clustering. These clusters are then color coded to reflect that.

In this article, we present a visualization based systematic domain analysis of the inter-disciplinary nature of agent-based computing. This involves the discovery of various types of co-citation networks as well as the complex network analysis of the overall network using two different tools based on their relative strengths; Network Work Bench (Pullen 2000; Team 2006) for performing a Complex Network Analysis using Network Analysis Toolkit and CiteSpace(Chen 2006).

The objectives of this study can be summarized as following:

- To identify the largest clusters of inter-connected papers for the discovery of complex interrelations of various sub-areas of agent-based computing.
- To discover "bursts"(Barabasi 2005; Barabási and Gelman 2010) of topics along with a timeline of publication of the index terms used.
- Identification of the core journals for the whole of agent-based computing ranging from agents, multi-agent systems to agent-based and individual-based simulations, in terms of citations of articles.
- Identification of the key subject categories.
- Identification and study of the most productive authors, institutes and countries of manuscript origins.

## Methodology

### Data collection

The input data was retrieved from the Thomson Reuters web of knowledge(Reuters 2008). A thorough topic search for data was devised to cater for various aspects and keywords used in agent-based computing in the following three sub-domains:

1. Agent-based, multi-agent based or individual-based models
2. Agent-oriented software engineering
3. Agents and multi-agent systems in AI

The search was performed in all four databases of Web of Science namely SCI-EXPANDED, SSCI, A&HCI, CPCI-S for all years. Details of the search have been provided in the Appendix. For the sake of analysis, the range of years 1990-2010 was selected with search limited to only Journal articles. Bibliographic records retrieved from SCI include information such as authors, titles, abstracts as well as references. The addition of cited references resulted in a total of 32, 576 nodes. The details of the search keywords along with a reasoning for the selection are given in the Appendix.

**Research methodology tools**

We chose two different tools for the analysis. The reason for this selection was two-fold; firstly that way we can capitalize on the strengths of each of these and secondly, this allows for cross-validation of results by comparison of the outputs from the two tools. The first tool was the Network WorkBench (Pullen 2000), which is considered a strong complex network analysis tool and was primarily used for global network analysis. The second was CiteSpace (Chen 2006), a recent tool which has been designed exclusively for Citation Networks Analysis. We used CiteSpace as the key visualization tool for our study.

Network workbench is a general purpose tool which is able to visualize, generate and analyze various types of complex networks. It has a specially developed interface for scientometric analysis(LaRowe et al. 2009; Börner et al. 2006). It is able to extract almost 8 types of citation networks from ISI text data. These include directed, paper citation, author-paper, co-occurrence, word co-occurrence, co-author, reference co-occurrence and document co-citation networks. Subsequently it can be used to perform analysis as well as for the visualization of networks. One key ability of Network Workbench is the fact that it can pre-process data (e.g. by reducing the number of nodes for analysis of important features) and can be used to manipulate networks using GUESS(Adar 2006).

CiteSpace, on the other hand, is custom designed for Citation Analysis visualization. By color coding the evolution of research, it allows examination of some details which cannot otherwise be easily captured using other tools. As an example, as shown by Chen in (Chen 2006), using color-coded time slicing, it can be used to identify papers which tie in two different research fronts.

# Results

We start with a basic look at the overall picture of articles retrieved from the Web of Science. As can be seen in Figure 1, the articles in this domain start primarily during the early 1990's and gradually keep rising and as such reach a total of 148 articles published merely in the last year 2009.

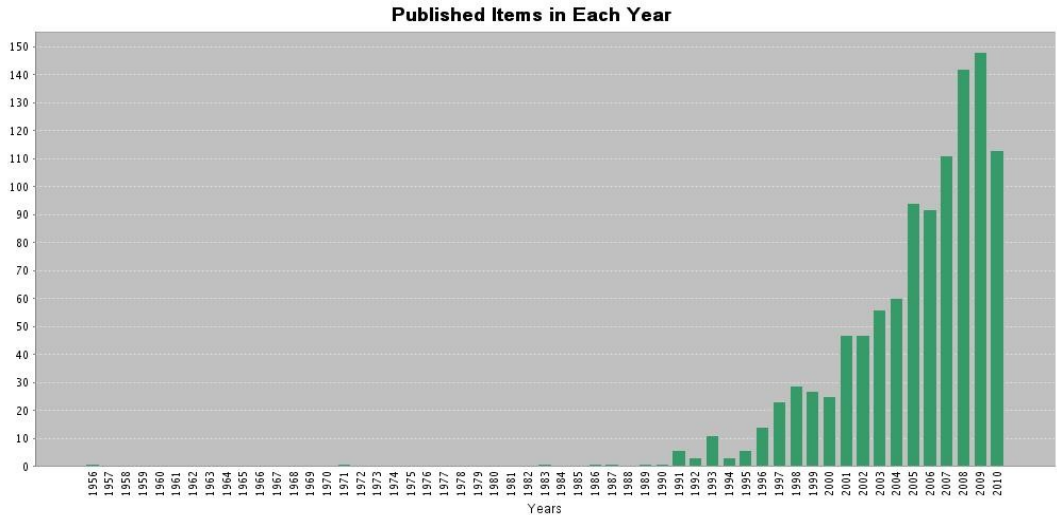

**Figure 1 Articles published yearly**

In addition, since the popularity of a domain is known to be based on its number of citations, we need to observe this phenomena closely. It can be observed in the graph using data from the Web of Science in Figure 2. Thus, starting from a very small number of citations, Agent-based computing has risen to almost 1630 citations alone during the year 2009.

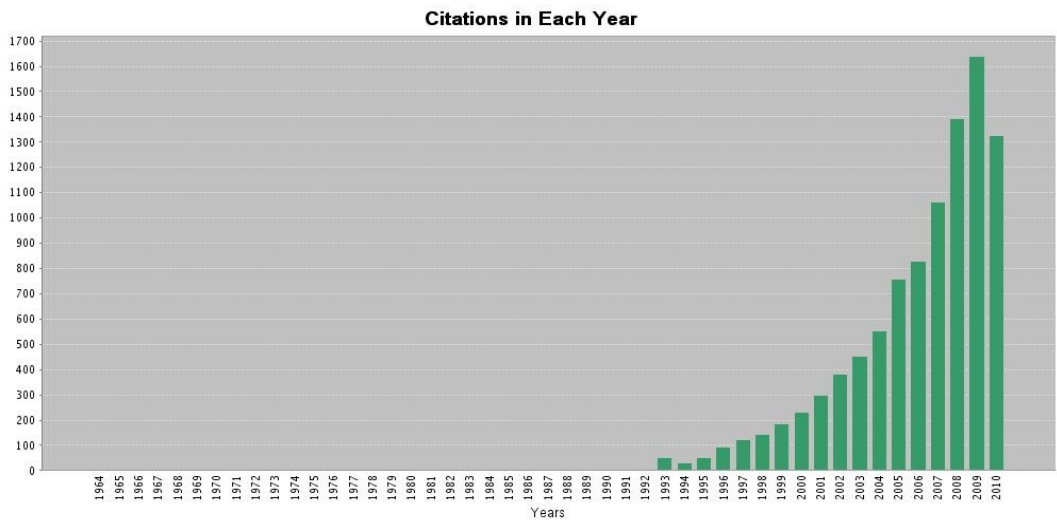

**Figure 2 Citations per year**

**Analysis using NWB**

Next, to get the big picture of the citation network, a network (paper co-citation network) was extracted from the ISI records using Network Workbench tool. In addition, analysis was performed for the extracted network using NWB based Network Analysis Toolkit (Sedgewick and Schidlowsky 2003). The extracted global properties of the network are shown in Table 1:

**Table 1 Basic Network Analysis of Extracted Paper Citation Network**

| Attribute | Value |
|---|---|
| Nodes: | 32, 576 |
| Isolated nodes: | None |
| Edges: | 39, 096 |
| Self Loops | None |
| Parallel Edges | None |
| Edge Attributes | None |
| Valued Network | No |
| Average total degree: | 2.400 |
| Weakly Connected | No |
| Weakly connected components | 78 |
| Largest connected component Size | 31, 104 nodes |
| Density (disregarding weights): | 0.00004 |

The first thing to note here is the number of nodes. The number of nodes is 32, 576 and the reason it is much larger than reported above is that the Web of Science data includes the cited references as well as the original nodes. Now, we see here that there are no isolated nodes, which is obvious because every paper will at least cite some other papers at the very least. This is followed by the 39, 096 edges. Another interesting feature is the average degree, which we see is 2.4. The fact that the average degree is not significantly higher shows actually that a large number of papers have been co-cited in this corpus otherwise the degree would have been much higher.

The graph itself is not weakly connected but it consists of 78 weakly connected components with a largest component of size 31,104 nodes. The weakly connected components are formed based on articles connected with a small number of other articles identifying the importance of these papers (Which shall be investigated in depth later on in this paper). Finally, we note that the density is 0.00004 however knowing the density does not give much structural information about the Scientific domain per se. These properties are a characteristic of the retrieved empirical data. However, we note here that while interesting, they do not provide in depth insights of the data. So we perform a further set of analysis using NWB

Here we examine the structure of the overall agent-based computing domain further using NWB. To allow for the examination of the domain using the specific strengths, central to the NWB tool, we extracted the top nodes with an arbitrary number of local citation count > 20. Subsequently the network was visualized using GUESS(Adar 2006) and nodes were resized according to the values of the citation count. The result can be observed in Figure 3. Here we can note the peculiar relation of some of the top papers connected with Volker Grimm's paper in the center. Apart from Grimm's own papers, these include Robert Axelrod's as well as Parker's and Deangelis's works.

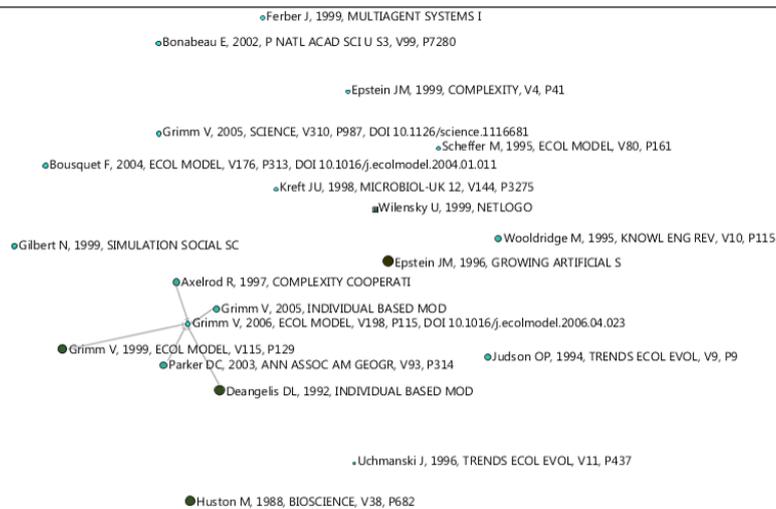

Figure 3 Top papers with citations > 20

In addition, we can note that here Kreft's and Huston's work in Microbiology and Biosciences as well as Wooldridge's and Ferber's work in the domain of multiagent systems are also showing up. For futher analysis, we move on to the more advanced Information Visualization tool, namely CiteScape.

**CiteScape**

The CiteScape tool allows for a variety of different analysis. CiteScape directly operates on the downloaded ISI data and builds various networks using time slicing. Subsequently using the various options selected by the user, the network can then be viewed in different ways and parameters can be analyzed based on centrality (betweenness) as well as frequency.

*Identification of the largest cluster*

The goal of our first analysis was to observe the big picture and identify the most important indexing terms. Based on a time slice of one year, here in Figure 4, we see the largest cluster.

These clusters are formed using CiteSpace, which analyzes clusters based on time slices. Here the links between items show the particular time slices.

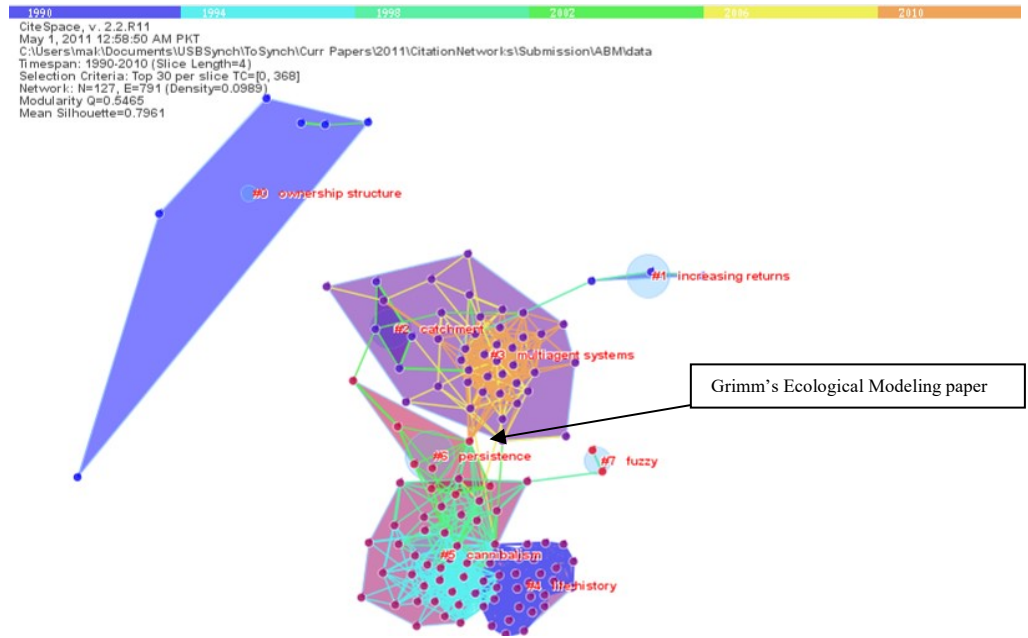

**Figure 4 Largest Cluster based on indexing terms**

In this figure, we first note the top where the year slices from 1990 to 2010 show up in 4 year slices. Using different colors in CiteSpace allows us to clearly identify turning points and other interesting phenomena in literature. We notice here that one of Grimm's papers is actually the key turning point from Agent-based Modeling to Multiagent-systems cluster. Another thing that can be noted here is the role of this paper in connecting a large number of papers in 2010 to the other papers in 2002-2006 era showing a somewhat delayed acceptance of the concepts.

*Timeline and bursts of index terms*

The next analysis performed was to observe the clusters on a timeline as shown in Figure 5.

In complex networks, various types of centrality measures such as degree centrality, eccentricity, closeness and shortest path betweenness centralities etc.

Citespace, in particular, uses betweenness centrality(Chen 2006). This particular centrality is known to note the ability of a vertex to monitor communication between other vertices. In other words, a higher centrality ensures that the vertex is between more of the shortest paths between other nodes relative to other nodes with lower centrality.

Using a time line especially helps identify the growth of the field. Please note here that these includes papers which are based on agent-based computing as well as papers which are cited by these papers. Here, the red items are the "bursts". Bursts identify sudden interest in a domain exhibited by the number of citations.

We can see that there are a lot of bursts in the domain of agent-based model. In addition, even our preliminary analytics and visualization here confirms the hypothesis that agent-based computing has a very wide spread article base across numerous sub-domains. This is obvious here as we see clusters ranging from "drosophila" and "yellow perch bioenergetics" to those based on "group decision support systems", all ranging from different domains. Further analysis in the paper strengthens this initial point of view based on an examination of the various clusters.

Here, the results demonstrate the effects of the semantic vagueness discussion earlier in the paper. So, e.g. Where concepts such as group decision support system, rule-based reasoning and coordination are concepts tied closely with developing intelligent agents, they also show up in the domain right along side agent-based modeling.

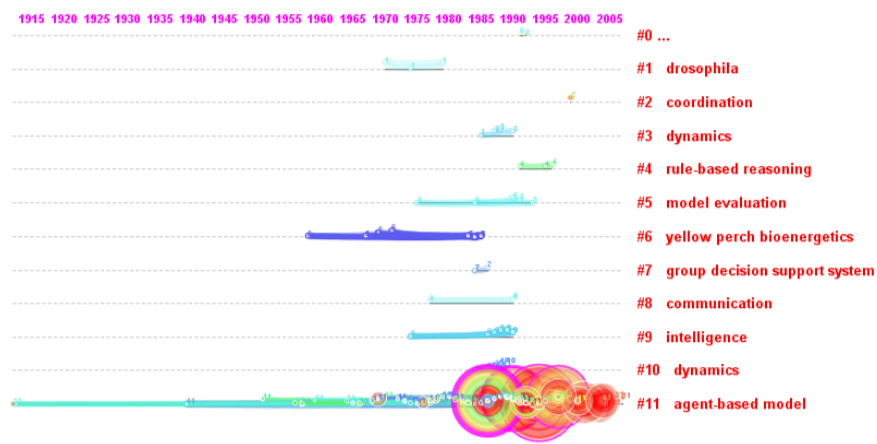

**Figure 5 Timeline view of terms and clusters of index terms (based on centrality) also showing citation bursts**

*Analysis of Journals*

Our next analysis was to identify the key publications of the domain[2]. This can be seen in Figure 6. Here the key journals are identified based on their centrality.

Once again, we can note here that the vagueness in the terms of use again shows up in the set of mainstream Journals of the domain with "Artificial Intelligence" and "Communications of the ACM" being relevant mostly to Agents associated with the concepts of "Intelligent agents" and "Multiagent Systems" whereas "Econometrica", "Ecological Modeling" and the journal "Science" representing the "agent-based modeling" perspective.

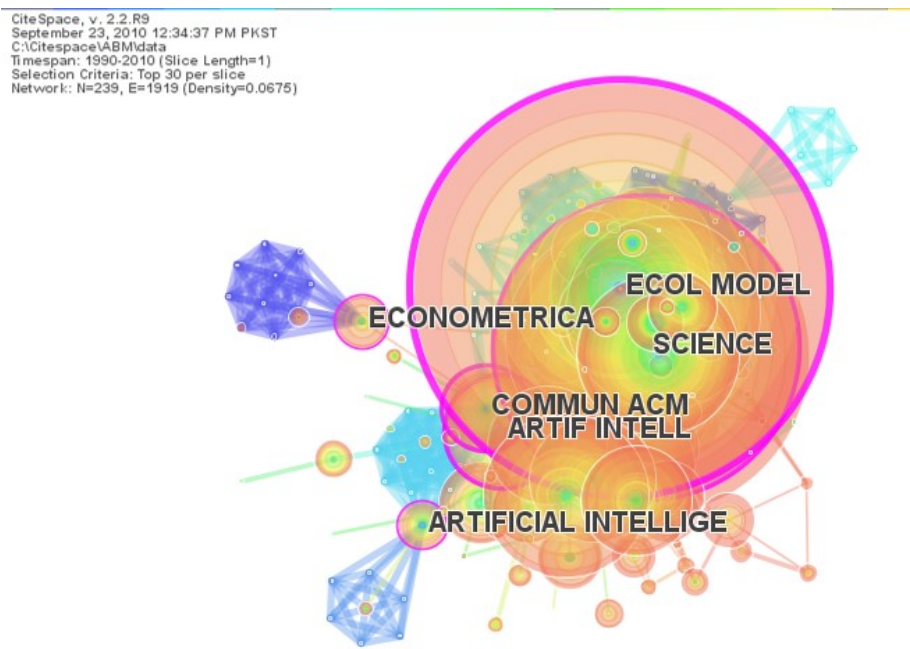

**Figure 6 Key Journals based on centrality**

In the following Table 2, we give details of these top journals based on centrality shown in the figure.

**Table 2 Top Journals based on Centrality**

| Centrality | Title | Abbreviated Title |
| --- | --- | --- |

---

[2] It is pertinent to note here that we faced one peculiar problem in the analysis of the retrieved ISI data. The Web of Science data identified a Journal named "Individual-based model". However extensive searches online did not find any such journal.

| | | |
|---|---|---|
| 0.47 | Ecological Modelling | ECOL MODEL |
| 0.29 | Science | SCIENCE |
| 0.21 | Communications of the ACM | COMMUN ACM |
| 0.32 | Artificial Intelligence | ARTIF INTELL and ARTIFICIAL INTELLIGE |
| 0.15 | Econometrica | ECONOMETRICA |
| 0.09 | Journal of Theoretical Biology | J THEOR BIOL |
| 0.08 | Canadian Journal of Fisheries and Aquatic Sciences | CAN J FISH AQUAT SCI |
| 0.08 | Nature | NATURE |
| 0.08 | Annals of the New York Academy of Sciences | ANN NY ACAD SCI |

The table above represents the centrality of the top ten key journals. In terms of centrality, the "ECOL MODEL" Journal has the highest value of centrality among all the journals. In addition, here we observe that "ANN NY ACAD SCI", "CAN J FISH AQUAT SCI", "NATURE" and the "ANN NY ACAD SCI" are also some of the top Journals of this domain in terms of Centrality.

Next, we analyze the publications in terms of their frequencies of publication as given in Table 3 below. Now, interestingly, the table sorted in terms of article frequency, gives a slightly different set of core journals. Through frequency analysis of the title words of 240 journals, it can be seen that "ECOL MODEL" is still at the top with a frequency of 295 articles. "NATURE" and "SCIENCE" follow with 231 and 216 published articles respectively. "J THEO BIO" is next with 167 articles. Next "ECOLOGY" has published 145 articles. This is followed by "AM NAT", "TRENDS ECOL EVOL", "LECT NOTES ARTIF INT" and "P NATL ACAD SCI USA".

**Table 3 Core Journals based on frequency**

| Frequency | Title | Abbreviated Title |
|---|---|---|

| 295 | Ecological Modelling | ECOL MODEL |
| 231 | Nature | NATURE |
| 216 | Science | SCIENCE |
| 167 | Journal of Theoretical Biology | J THEOR BIOL |
| 145 | Ecology | ECOLOGY |
| 123 | The American Naturalist | AM NAT |
| 121 | Trends in Ecology & Evolution | TRENDS ECOL EVOL |
| 121 | Lecture Notes in Artificial Intelligence | LECT NOTES ARTIF INT |
| 121 | Proceedings of the National Academy of Sciences | P NATL ACAD SCI USA |

*Analysis of Categories*

Our next analysis was to discover the prevalence of various agent-based computing articles in various subject categories. This visualization is shown in Figure 7. The detailed analysis of the subject category based on centrality follows in Table 4.

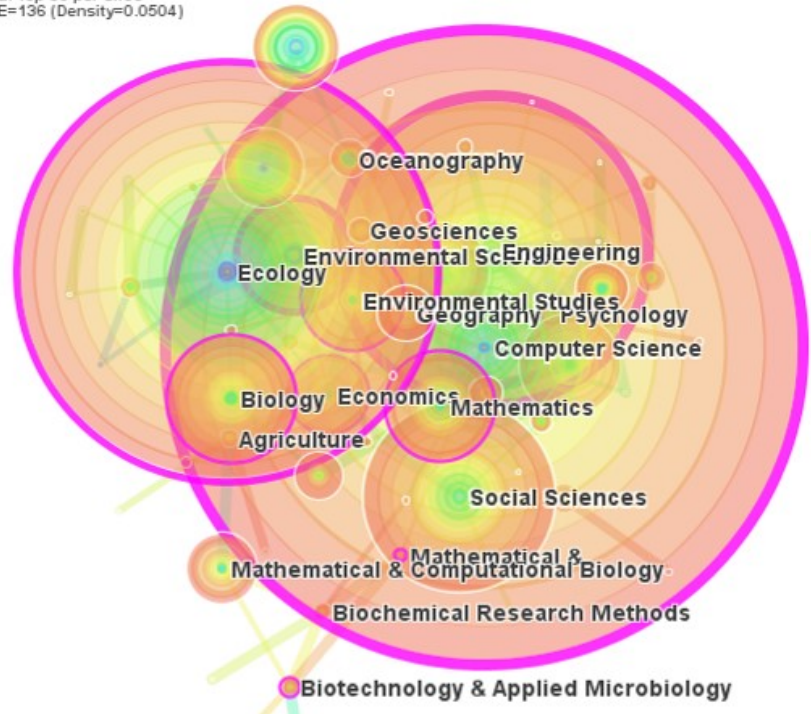

Figure 7 Category data

Table 4 Key categories based on Centrality

| Centrality | Category |
|---|---|
| 0.44 | Engineering |
| 0.4 | Computer Science |
| 0.36 | Mathematics[3] |
| 0.34 | Ecology |
| 0.31 | Environmental Sciences |
| 0.14 | Biotechnology & Applied Microbiology |
| 0.13 | Biology |
| 0.13 | Economics |
| 0.12 | Psychology |

The above table represents the centrality based ordering of the key subject categories. It is important to note here is that this table shows top categories from a total of 75 categories. Here, it can be observed that in terms of centrality, the "Engineering" category leads other categories. It is however, closely followed by Computer Science, Mathematics, Ecology and Environmental Sciences. It appears however that the "Psychology" category has the lowest value of centrality among all other categories.

---

[3] Shown as two categories erroneously by CiteScape

For comparative analysis, we also analyze these categories using the publication frequency of articles. The results of this analysis are presented below in Table 5.

**Table 5 Subject Categories according to frequency**

| Frequency | Category |
|---|---|
| 287 | Computer Science |
| 195 | Ecology |
| 145 | Engineering |
| 92 | Social Sciences |
| 66 | Biology |
| 57 | Environmental Sciences |
| 57 | Mathematics |
| 53 | Environmental Studies |
| 52 | Operations Research & Management Science |
| 52 | Fisheries |

The table represents the frequency of the top ten key categories. Through frequency analysis of the title words of 75 categories, we interestingly come up with a slightly different set of results. Here, "Computer Science" with a frequency of 287 articles leads the rest and is followed closely by Ecology and Engineering. An interesting observation based on the two tables is that there are certain categories which have relatively low frequency but are still central (in terms of having more citations) such as Mathematics. Amongst the top categories, we can also see categories with a relatively lower frequency such as 53 for "Environmental Sciences" and 52 for "Operations Research & Management Science" as well as "Fisheries". This detailed analysis shows that prevalence of agent-based computing is not based on a few sporadic articles in a variety of subject categories. Instead, there are well-established journals and researchers with interest in and publishing a considerable number of papers in this domain, especially agent-based modeling and simulation.

*Bursts in subject categories*

Next, we analyze how various subject categories have exhibited bursts. This is shown in the Table 6 below. Here we can see that fisheries have the largest burst associated with the year 1991. Next are two closely related categories "Marine and Freshwater Biology" and "Ecology" in the same time frame. One very interesting finding here is that there are a lot of bursts in non-computational categories.

Table 6 Key bursts in subject categories

| Burst | Category | Year |
| --- | --- | --- |
| 13.09 | Fisheries | 1991 |
| 10.3 | Marine & Freshwater Biology | 1991 |
| 9.36 | Ecology | 1991 |
| 5.58 | Economics | 1996 |
| 4.25 | Evolutionary Biology | 1993 |
| 3.76 | Mathematics | 1990 |
| 3.3 | Genetics & Heredity | 1993 |
| 3.2 | Oceanography | 1995 |

**Analysis of Author Networks**

In this section, we analyze the author co-citation networks. Figure 8 shows the visualization of the core authors of this domain. Here red color exhibits burst of articles and concentric circles identify separation of years of publications. The size of the circles is an indicator of the centrality of the author. Here blue is for older papers, green is for middle years whereas yellowish and reddish colors are for relatively more recent publications. Here, the descriptions are based on the color figures, which can be viewed in the online version of the paper, since Citespace, being a visualization tool, relies extensively on the use of colors to depict styles.

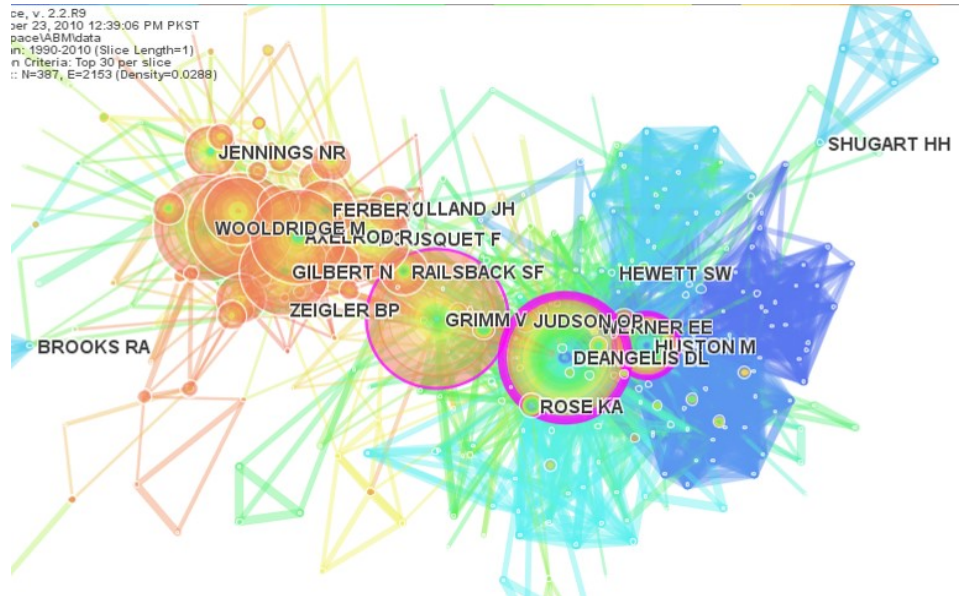

**Figure 8 Co-Author network visualization**

Although this visualization perhaps gives a broad picture of the various authors, we also present a detailed analysis of this data. This can be seen in a tabular form as shown below in Table 7. Here, we can observe that the top cited (most central) author is Don DeAngelis, a Biologist. Don is followed by another Biologist Michael Huston. Next is Volker Grimm, an expert in agent-based and individual-based modeling and Kenneth A Rose, an ecologist. Next, we have Robert Axelrod, a Political Scientist. Nigel Gilbert, a Sociologist and Mike Wooldridge, a Computer Scientist is next in the list. Finally, we have François Bousquet from the field of Ecological Modeling (Agriculture) and Steven F. Railsback, an Ecologist. This is quite an interesting result because agents and agent-based computing in general was supposed to be primarily from Computer Science/AI and have very specific meanings. However, our results show that it is actually quite prevalent and flourishing in an uninhibited manner in various other fields in terms of renowned (ISI-indexed) archival Journal articles.

**Table 7 Authors in terms of centrality**

| Centrality | Author |
|---|---|
| 0.5 | DEANGELIS DL |
| 0.33 | HUSTON M |

| 0.16 | GRIMM V |
| 0.08 | ROSE KA |
| 0.08 | AXELROD R |
| 0.07 | GILBERT N |
| 0.07 | WOOLDRIDGE M |
| 0.06 | BOUSQUET F |
| 0.06 | RAILSBACK SF |

For further comparative analysis, we plotted the top authors in terms of frequency of publications. This is shown in Figure 9.

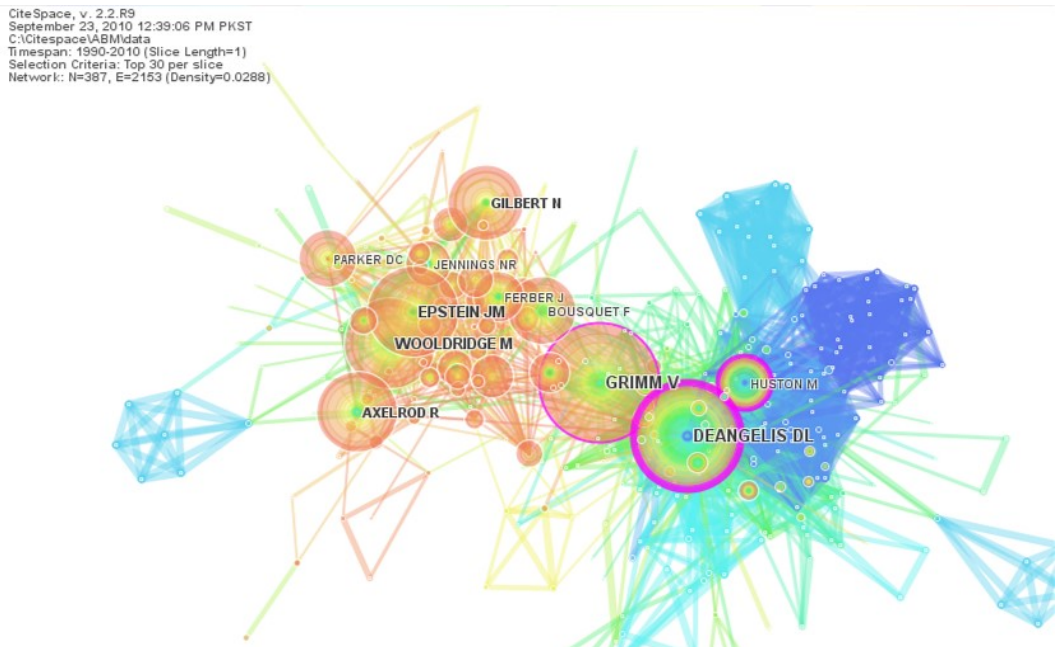

**Figure 9 Authors in terms of frequency**

The detailed analysis of this is shown below in tabular form of Table 8 below for these authors and their relative frequency of publication.

**Table 8 Top authors based on frequency**

| Frequency | Author |
| 128 | GRIMM V |
| 118 | DEANGELIS DL |
| 101 | EPSTEIN JM |
| 99 | WOOLDRIDGE M |
| 90 | AXELROD R |
| 83 | GILBERT N |
| 76 | BOUSQUET F |
| 68 | HUSTON M |
| 64 | FERBER J |
| 59 | PARKER DC |

It is important to note here that the results are based on a total of 387 cited authors. New names which appear in this table include Joshua M. Epstein, a Professor of Emergency Medicine. Prof. Epstein is an expert in human behavior and disease spread modeling. Jacques Ferber, a Computer Scientist is another new name in the list along with Dawn C. Parker, an agricultural Economist.

**Country-wise Distribution**

Next, we present an analysis of the spread of research in agent-based computing originating from different countries based on centrality. Here, in Figure 10, we can see the visualization of various countries. Please note here that the concentric circles of different colors/shades here represent papers in various time slices (we have selected one year as one time slice). The diameter of the largest circle thus represents the centrality of the country. Thus the visualization identifies the key publications in the domain to have originated from the United States of America. This is followed by papers originating from countries such as England, China, Germany, France and Canada.

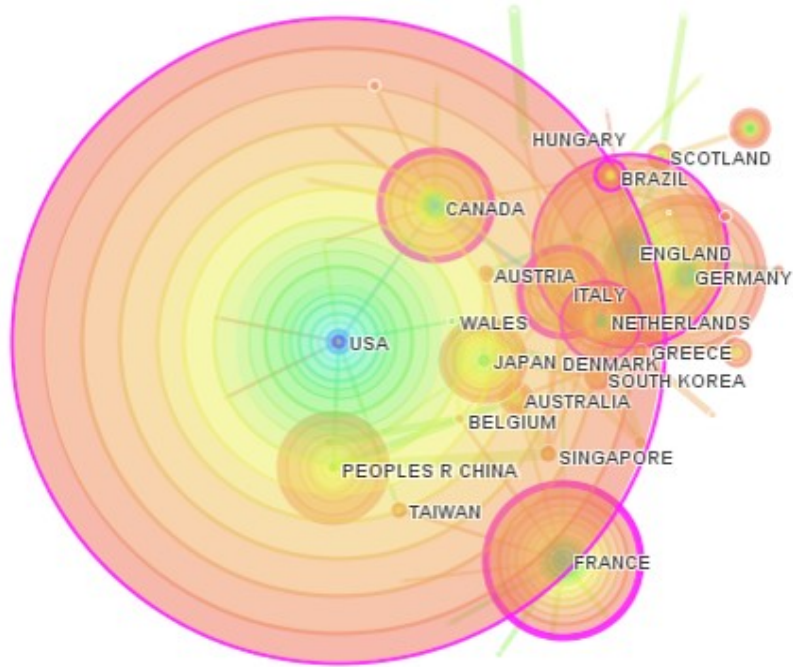

**Figure 10 Countries with respect to centrality**

## Analysis of Institutes

In this sub-section, we present visualization for a detailed analysis of the role of various Institutes. We can see the temporal visualization of various institutes in the domain and the assortment of popular keywords associated with them on the right in Figure 11.

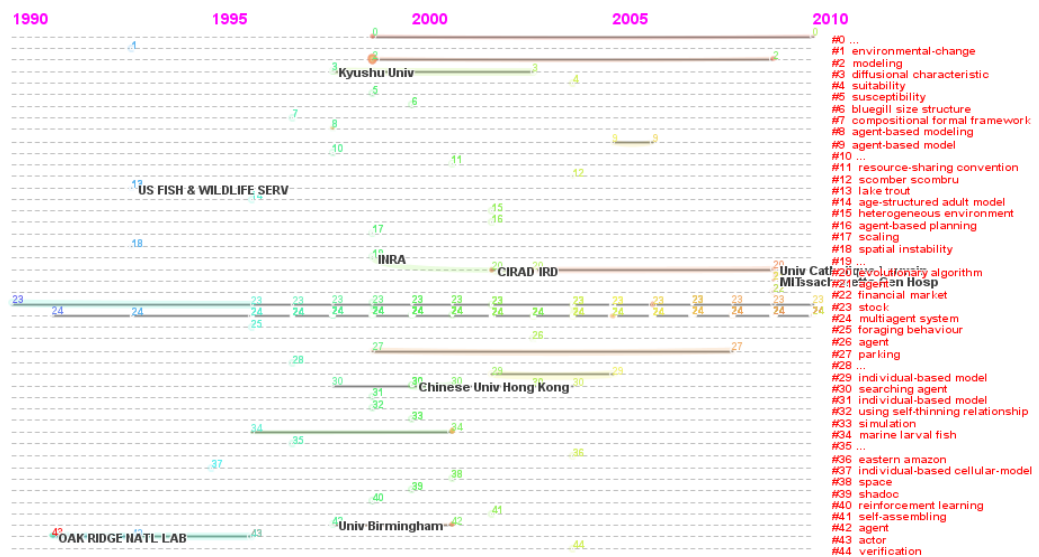

**Figure 11 Institutes**

Here, it can be observed that the prevalence of manuscripts originating from the Oak Ridge National Laboratories from the early 1990's. Next, we see papers from US Fish and Wildlife services. Then, we can see papers from Kyushu

University, Japan, University of Birmingham, UK and French National Institute of Agricultural Research (INRA) close to 1998. From this time to 2000, a prevalent institute is Chinese University, Hong Kong. In around 2002, the French institute Centre de coopération Internationale en recherche agronomique pour le développement (CIRAD), an Institute of Agriculture Research for Development is prevalent followed closely by a sister institute, the Institut de Recherche pour le Développement (IRD). More recent newcomers to the field of agents include the MIT and the Massachusetts General Hospital, associated with the Harvard University.

In the following table, we perform an alternative analysis which is based instead on the frequency of articles.

**Table 9 Core Institutes based on frequency**

| Frequency | Institute |
|---|---|
| 11 | University of Illinois, USA |
| 10 | INRA, France |
| 9 | University of Michigan, USA |
| 9 | University of Minnesota, USA |
| 8 | University of Sheffield, UK |
| 8 | Nanyang Technological University, Singapore |
| 8 | Italian National Research Council (CNR), Italy |
| 8 | Oak Ridge National Labs, USA |
| 7 | Harvard University, USA |
| 7 | MIT, USA |
| 7 | University of Washington, USA |
| 7 | University of Hong Kong, P. R. China |
| 7 | IRD, France |

The Table 9 listed above represents the top institutes in terms of frequency. It should be noted that the frequency analysis was based on title words of a total of 328 institutes. "University of Illinois"[4] in general has a top ranking with a frequency of 11 articles. It is followed closely by INRA (France) with a frequency of 10 articles. University of Michigan and University of Minnesota, both from the US follow next with 9 articles each. With 8 articles each, next we have University of Sheffield (UK), Nanyang Technological University (Singapore), Italian National Research Council (CNR) and the Oak Ridge National Labs (USA).

**Summary of results**

In the current paper, we have used two different pieces of software to analyze various types of visualization and results emanating from the agent-based computing domain. In this subsection, we want to emphasize the various results that have been previously identified inline with the theme of agent-based computing.

---

[4] ISI data extracted using CiteSpace does not differentiate further as to which exact campus of the University of Illinois is considered here (of the primarily three campuses i.e. UIUC, UIC, UIS.)

Firstly, we noted in the analysis using Network Workbench that the top cited papers of the domain contained papers from both the multiagent side of the domain (based in Computer Sciences) in addition to the agent-based modeling ideas (from Ecology/Social sciences). Next, using clustering of index terms, we observed that the originating ideas actually came from agent-based modeling. Whereas, we observed a major recent turning point with some of Grimm's papers. However, another yet older turning point was Goldberg's 1989 paper on Genetic Algorithms, which tied in the Ecological cluster labeled as "Cannibalism". In addition, we see prevalence of the "catchment" cluster identifying the social sciences and the "increasing returns" cluster identifying economics. This single figure shows how these diverse areas are tied in together and are linked by citing documents. The "persistence" cluster again identifies a Computer Science "multiagent" related set of documents. Here, the oldest cluster is the "life history" cluster with roots in Biosciences. This strong intermingling of ideas has silently lead to the evolution of agent-based computing domain. Whereas subsequent analyses only prove the point that there is further diversity in each of subject categories, authors, institutions and countries. However, while the domain has strong footing, being able to use visualization has thus allowed the examination of a complex interaction of sub-domains, which would not have been possible without the use of these tools and techniques. It is important for the practitioners and researchers working in the sub-domains because firstly it clarified the reason for the confusion in the usage terms. Secondly, it allows researchers to identify new research fronts in parallel domains, which can be re-used in other domains to minimize re-inventing the wheel.

## Conclusions and Future Work

In this paper, we have presented a detailed visual and scientometric survey of the entire agent-based computing domain, covering all Journal articles in Thomson Reuters from the era 1990-2010 for papers ranging from diverse domains such as agent-based modeling and simulation, agent-based software engineering, multi-agent systems and artificial intelligence. Our approach in this survey was based on actual data from the recognized Web of Science databases. This allowed us to cover all articles in the set of seemingly diverse domains coming under agent-based computing. Our analysis has produced some interesting results. One such result is that contrary to what can expected about the evolution of the domain, with roots in the Computer Sciences, we have discovered that the domain is also quite well-established and actually has a relatively higher number of cited articles in a number of non-computing domains. These range from the life sciences to ecological sciences and even the social sciences. In the future, we plan on performing a detailed analysis of each of these sub-domains taken individually to further highlight the scientometric indicators inside each of multi-agent systems, agent-oriented software engineering and agent-based modeling. We are also currently working on performing visualization based analyses of somewhat lesser- related areas such as "Cybernetics" and even "Consumer Electronics".

## References

Adar, E. Guess: a language and interface for graph exploration. In*, 2006* (pp. 800): ACM

Aoyama, H. (2010). *Econophysics and companies : statistical life and death in complex business networks*. New York: Cambridge University Press.

Axelrod, R. (1997). *The complexity of cooperation: agent-based models of competition and colloboration.* Princeton, NJ: Princeton University Press.

Bankes, S. C. (2002). Agent-based modeling: A revolution? (Vol. 99, pp. 7199-7200): National Acad Sciences.

Barabasi, A. (2005). The origin of bursts and heavy tails in human dynamics. *Nature, 435*(7039), 207-211.

Barabási, A., & Gelman, A. (2010). Bursts: The Hidden Pattern Behind Everything We Do. *Physics Today, 63*, 46.

Barabási, A., Jeong, H., Néda, Z., Ravasz, E., Schubert, A., & Vicsek, T. (2002). Evolution of the social network of scientific collaborations. *Physica A: Statistical Mechanics and its Applications, 311*(3-4), 590-614.

Bellifemine, F., Poggi, A., & Rimassa, G. (2001). Developing multi-agent systems with JADE. *Intelligent Agents VII Agent Theories Architectures and Languages*, 42-47.

Börner, K., Penumarthy, S., Meiss, M., & Ke, W. (2006). Mapping the diffusion of scholarly knowledge among major US research institutions. *Scientometrics, 68*(3), 415-426.

Card, S., Mackinlay, J., & Shneiderman, B. (1999). *Readings in information visualization: using vision to think*: Morgan Kaufmann.

Chen, C. (2006). CiteSpace II: Detecting and visualizing emerging trends and transient patterns in scientific literature. *Journal of the American Society for Information Science and Technology, 57*(3), 359-377.

Chen, C., Panjwani, G., Proctor, J., Allendoerfer, K., Aluker, S., Sturtz, D., et al. (2006). Visualizing the Evolution of HCI. *People and Computers XIX— The Bigger Picture*, 233-250.

Chen, C., Paul, R., & O'Keefe, B. (2001). Fitting the jigsaw of citation: Information visualization in domain analysis. *Journal of the American Society for Information Science and Technology, 52*(4), 315-330.

Chun-juan, L., Yue, C., & Hai-yan, H. (2010). Scientometric & Visualization Analysis of Innovation Studies International. *Technology and Innovation Management, 1*.

Dydenko, I., Durning, B., Jamal, F., Cachard, C., & Friboulet, D. (2005). An autoregressive model-based method for contrast agent detection in ultrasound radiofrequency images. *Ultrason Imaging, 27*(1), 37-53.

Epstein, J., & Axtell, R. (1996). *Growing artificial societies*: MIT press Cambridge, MA.

Etzkowitz, H., & Leydesdorff, L. (2000). The dynamics of innovation: from National Systems and. *Research policy, 29*(2), 109-123.

Gilbert, N., & Troitzsch, K. G. (2005). *Simulation for the social Scientist* (Second ed.): McGraw Hill Education.

Gonzalez, P. P., Cardenas, M., Camacho, D., Franyuti, A., Rosas, O., & Lagunez-Otero, J. (2003). Cellulat: an agent-based intracellular signalling model. *Biosystems, 68*(2-3), 171-185, doi:S0303264702000941 [pii].

Griffith, B., Small, H., Stonehill, J., & Dey, S. (1974). The structure of scientific literatures II: Toward a macro-and microstructure for science. *Science studies, 4*(4), 339-365.

Hou, H., Kretschmer, H., & Liu, Z. (2008). The structure of scientific collaboration networks in Scientometrics. *Scientometrics, 75*(2), 189-202, doi:10.1007/s11192-007-1771-3.

Jennings, N. (1999a). *Agent-based computing: Promise and perils.* Paper presented at the 16th Int. Joint Conf. on Artificial Intelligence (IJCAI-99), Stockholm, Sweden,

Jennings, N. Agent-based computing: Promise and perils. In*, 1999b* (Vol. 16, pp. 1429-1436): Citeseer

LaRowe, G., Ambre, S., Burgoon, J., Ke, W., & Börner, K. (2009). The Scholarly Database and its utility for scientometrics research. *Scientometrics, 79*(2), 219-234.

Lesser, S. A. a. V. Multiagent Reinforcement Learning and Self-Organization in a Network of Agents. In *AAMAS 07, Honolulu, Hawaii, 2007*: ACM

Leydesdorff, L. (2001). *The challenge of scientometrics: The development, measurement, and self-organization of scientific communications*: Universal-Publishers.

Macal, C. M., & North, M. J. (2007). *Agent-based modeling and simulation: desktop ABMS.* Paper presented at the Proceedings of the 39th conference on Winter simulation: 40 years! The best is yet to come, Washington D.C.,

Mukhopadhyay, R., Costes, S. V., Bazarov, A. V., Hines, W. C., Barcellos-Hoff, M. H., & Yaswen, P. (2010). Promotion of variant human mammary epithelial cell outgrowth by ionizing radiation: an agent-based model supported by in vitro studies. *Breast Cancer Res, 12*(1), R11, doi:bcr2477 [pii] 10.1186/bcr2477.

Niazi, M., & Hussain, A. (2009). Agent based Tools for Modeling and Simulation of Self-Organization in Peer-to-Peer, Ad-Hoc and other Complex Networks. *IEEE Communications Magazine, 47*(3), 163 - 173.

Niazi, M., & Hussain, A. (2010). A Novel Agent-Based Simulation Framework for Sensing in Complex Adaptive Environments (In-Press). *IEEE Sensors Journal, 11*(0), doi:10.1109/JSEN.2010.2068044.

Niazi, M., & Hussain, A. (2011). *Social Network Analysis Of Trends In The Consumer Electronics Domain.* Paper presented at the International

Conference on Consumer Electronics, Las Vegas, USA, 9-12 January, 2011

Niazi, M., Siddique, Q., Hussain, A., & Kolberg, M. (2010). *Verification and Validation of an Agent-Based Forest Fire Simulation Model.* Paper presented at the SCS Spring Simulation Conference, Orlando, FL, USA, April 2010

Niazi, M. A., & Hussain, A. (2011). A Novel Agent-Based Simulation Framework for Sensing in Complex Adaptive Environments. *Sensors Journal, IEEE, 11*(2), 404-412.

Niazi, M. A., & Hussain, A. (2011). Sensing Emergence in Complex Systems. *IEEE Sensors Journal*, doi:10.1109/JSEN.2011.2142303.

Park, H., Hong, H., & Leydesdorff, L. (2005). A comparison of the knowledge-based innovation systems in the economies of South Korea and the Netherlands using Triple Helix indicators. *Scientometrics, 65*(1), 3-27.

Park, H., & Leydesdorff, L. (2008). Korean journals in the Science Citation Index: What do they reveal about the intellectual structure of S&T in Korea? *Scientometrics, 75*(3), 439-462.

Pouris, A., & Pouris, A. (2010). Scientometrics of a pandemic: HIV/AIDS research in South Africa and the World. *Scientometrics*, 1-12, doi:10.1007/s11192-010-0277-6.

Pullen, M. (2000). The Network Workbench: network simulation software for academic investigation of Internet concepts. *Computer Networks, 32*(3), 365-378.

Reuters, T. (2008). Web of science. *Online factsheet Thomson Reuters, Philadelphia, Pennsylvania (Available from: [www.thomsonreuters.com/content/PDF/scientific/Web_of_Science_factsheet.pdf](www.thomsonreuters.com/content/PDF/scientific/Web_of_Science_factsheet.pdf) )*.

Sandström, E., & Sandström, U. CiteSpace Visualization of Proximity Clusters in Dentistry Research. In*, 2007* (pp. 25–28)

Sedgewick, R., & Schidlowsky, M. (2003). *Algorithms in Java, Part 5: Graph Algorithms*: Addison-Wesley Longman Publishing Co., Inc. Boston, MA, USA.

Siddiqa, A., Niazi, M. A., Mustafa, F., Bokhari, H., Hussain, A., Akram, N., et al. A new hybrid agent-based modeling & simulation decision support system for breast cancer data analysis. In *Information and Communication Technologies, 2009. ICICT '09. International Conference on, 15-16 Aug. 2009 2009* (pp. 134-139)

Sierra-Flores, M., Guzmán, M., Raga, A., & Pérez, I. (2009). The productivity of Mexican astronomers in the field of outflows from young stars. *Scientometrics, 81*(3), 765-777.

Small, H. (1973). Co citation in the scientific literature: A new measure of the relationship between two documents. *Journal of the American Society for Information Science, 24*(4), 265-269.

Small, H. (1993). Macro-level changes in the structure of co-citation clusters: 1983–1989. *Scientometrics, 26*(1), 5-20.

Small, H., & Griffith, B. (1974). The structure of scientific literatures I: Identifying and graphing specialties. *Science studies*, 17-40.

Snead, D., Barre, P., Bajpai, P. K., Taylor, A., Reynolds, D., Mehling, B., et al. (1995). The use of a zinc based bioceramic as an osteoconductive agent in the rat model. *Biomed Sci Instrum, 31*, 141-146.

Sooryamoorthy, R. (2010). Scientific publications of engineers in South Africa, 1975–2005. *Scientometrics*, 1-16, doi:10.1007/s11192-010-0288-3.

Team, N. (2006). Network Workbench Tool. Indiana University, Northeastern University, and University of Michigan.

Weiss, G. (Ed.). (1998). *Multiagent Systems, A Modern Approach to Distributed Artificial Intelligence*. Cambridge, Massachusetts: The MIT Press.

White, H., & Griffith, B. (1981). Author cocitation: A literature measure of intellectual structure. *Journal of the American Society for Information Science, 32*(3), 163-171.

White, H., & McCain, K. (1998). Visualizing a discipline: An author co-citation analysis of information science, 1972-1995. *Journal of the American Society for Information Science, 49*(4), 327-355.

White, H. D., & McCain, K. W. (1998). Visualizing a discipline: An author co-citation analysis of information science, 1972–1995. *Journal of the American Society for Information Science, 49*(4), 327-355, doi:10.1002/(sici)1097-4571(19980401)49:4<327::aid-asi4>3.0.co;2-4.

Wooldridge, M. (1998). Agent-based computing. *Interoperable Communication Networks, 1*, 71-98.

Wooldridge, M. (2009). *An introduction to multiagent systems*: Wiley.

Xiaofei, D. (2010). A Study on Simulation of Forest Fire Spread Based on a Multi-Agent Model. *Anhui Agricultural Science Bulletin*.

Zhao, R., & Wang, J. (2010). Visualizing the research on pervasive and ubiquitous computing. *Scientometrics*, 1-20.

# Appendix A: Details of Search Keywords

Here we would like to mention the keywords used for searching the ISI web of knowledge in addition to the reasoning behind the selection. Arguably there are numerous ways to classify as sub-domain based on keywords. In this particular case, some of the keywords were even shared with Chemical and Biological Journals (e.g. using the word agent for e.g. Biological agent or Chemical agent even). As such, we had to limit the search to papers with a focus on either agent-based modeling specifically or else in the domain of multiagent systems.

The search was thus performed on titles and the exact search from the ISI web of knowledge was as following:

Title=(agent-based OR individual-based OR multi-agent OR multiagent OR ABM*) AND Title=(model* OR simulat*)

Timespan=All Years. Databases=SCI-EXPANDED, SSCI, A&HCI, CPCI-S.

Date retrieved: 8$^{th}$ September 2010 (1064 records)